\documentclass[11pt]{iopart}
\usepackage{iopams}
\usepackage{enumerate}
\usepackage{amsfonts}
\usepackage{amssymb}
\usepackage{amsthm}
\usepackage{braket}
\usepackage{mathrsfs}
\usepackage{dcolumn}
\usepackage{bm}
\usepackage{graphicx, graphics}
\usepackage{color}
\usepackage[pdftex]{hyperref}
\hypersetup{
colorlinks=true,
linkcolor=[rgb]{0.0,0.1,0.8},
citecolor=[rgb]{1.0,0.0,0.0},
filecolor=magenta,
urlcolor=[rgb]{0.0,0.3,1}
}

\newcommand{\blue}[1]{{\color{blue} #1}}

\definecolor{codegreen}{rgb}{0,0.6,0}

\theoremstyle{Theorem}

\newcommand{\pap}[1]{\left( #1 \right)}
\newcommand{\pas}[1]{\left[#1 \right]}

\newcommand\restr[2]{{
\left.\kern-\nulldelimiterspace 
#1 
\vphantom{\big|} 
\right|_{#2} 
}}

\begin{document}
\title[Cavity Control of Topological Qubits]{Cavity Control of Topological Qubits: Fusion Rule, Anyon Braiding and Majorana-Schr\"odinger Cat States}

\author{Luis Quiroga$^1$\href{https://orcid.org/0000-0003-2235-3344}{\includegraphics[scale=0.5]{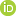}},
Fernando J. G\'omez-Ruiz$^{2,3}$\href{https://orcid.org/0000-0002-1855-0671}{\includegraphics[scale=0.5]{orcid}},\\
Ivan A. Bocanegra-Garay$^{3}$\href{https://orcid.org/0000-0002-5401-7778}{\includegraphics[scale=0.5]{orcid}},
Ferney J. Rodr\'iguez$^{1}$\href{https://orcid.org/0000-0001-5383-4218}{\includegraphics[scale=0.5]{orcid}},
and\\
Carlos Tejedor$^{4,5}$\href{https://orcid.org/0000-0001-5383-4218}{\includegraphics[scale=0.5]{orcid}}
}
\address{$^1$ Departamento de F{\'i}sica, Universidad de los Andes, A.A. 4976, Bogot\'a D.C, Colombia}

\address{$^2$ Departamento de F\'isica, Universidad Carlos III de Madrid, Avda. de la Universidad 30, 28911 Legan\'es, Spain}

\address{$^3$ Departamento de F\'isica Te\'orica, At\'omica y \'Optica and  Laboratory for Disruptive Interdisciplinary Science, Universidad de Valladolid, 47011 Valladolid, Spain}
\address{$^4$ Departamento de F{\'i}sica Te{\'o}rica de la Materia Condensada, Universidad Aut{\'o}noma de Madrid, Madrid 28049, Spain}
\address{$^5$ Condensed Matter Physics Center (IFIMAC), Universidad Aut{\'o}noma de Madrid, Madrid 28049, Spain}
\ead{\href{mailto:fegomezr@fis.uc3m.es}{fegomezr@fis.uc3m.es}}
\begin{abstract}

We investigate the effects of coupling a local electromagnetic cavity to a segment of a topological Kitaev chain (KC), with particular emphasis on the interplay between photons and Majorana zero modes (MZMs). In addition to the well-known {\it scissor effect}-which effectively partitions the chain and isolates free MZMs in the bulk-we provide evidence of non-trivial fusion rules and braiding operations, hallmark signatures of non-Abelian anyons, enabled by spatially selective ultrastrong KC-cavity coupling. We propose that these distinctive MZM properties can be experimentally probed via fermionic parity measurements and photon-induced Berry phases. Furthermore, we demonstrate that, in the so-called sweet-spot regime, the coupled system can be mapped onto a Rabi-like model with a homodyne-rotated quadrature, offering a simplified yet powerful theoretical description. Exploiting the symmetry of fermionic modes within a two-site cavity configuration, we also show the feasibility of generating hybrid MZM-polariton Schr\"odinger cat states. Our findings offer a novel approach to manipulating topological quantum matter through local light-matter interactions and provide theoretical tools for future experimental realizations in platforms such as quantum materials or superconducting circuits.
\end{abstract}
\\
\submitto{\href{https://iopscience.iop.org/article/10.1088/1402-4896/adf782}{Phys. Scr. {\bf 100}, 085122 (2025)}}
\maketitle

\section{Introduction}

Topological quantum computing is the holy grail of current research on quantum technologies~\cite{das}. Its key building blocks are non-Abelian anyon quasi-particles in topological matter~\cite{nayak, stern, song}. Clear experimental evidence on such anyons is still missing, but recent advances in precision control of hybrid semiconductor-superconductor nanowires and quantum dot arrays have brought researchers closer to their realization~\cite{marcus, Dvir2023}. This opens new possibilities in designing novel experimental frameworks while raising conceptual questions of great relevance.

The essential physics of 1D Ising anyons is captured by the Kitaev chain (KC) model~\cite{kitaev}. In recent years, quantum-dot (QD) superconductor arrays have emerged as a promising platform for realizing Kitaev chains~\cite{Sau2012,dourado2025}. Notably, a minimal two-site configuration has been successfully implemented in low-dimensional semiconductor systems~\cite{Leijnse_PRB12}, with tunneling spectroscopy providing direct evidence of MZM at a specific parameter regime~\cite{Dvir2023,tenHaaf2024,Zatelli2024,Bordi_PRL24}. These advancements position hybrid semiconductor-superconductor heterostructures as highly versatile platforms, driving significant progress in developing topological solid-state qubits~\cite{Aghaee_2025,miles2025}.  A particularly promising configuration is the so-called “sweet spot” regime, where fine-tuned experimental conditions highlight Majorana-like physics, even though the system may lack full topological protection.

In a KC, each local Dirac fermion of the chain can be expressed in terms of a pair of Majorana fermions (MF). A key consequence is the existence of a topological phase in which a zero-energy mode exists. It consists of two MF each one localized at one edge of the chain. Occupied and unoccupied options of such state have zero energy, forming a non-local qubit. Quantum information is encoded in the collective state of both Majoranas, and this nonlocal nature imparts redundancy and resilience to decoherence. Moreover, the zero-energy Majorana modes (MZMs) become decoupled from the rest of the KC, making them ideal candidates for non-Abelian statistics. Topological protection is produced through braiding and measurement of the non-Abelian anyon character of MZM.

Recent quantum simulation experiments have significantly advanced our understanding of braiding and quantum teleportation using delocalized Majorana qubits, particularly in Kitaev chains coupled to superconducting resonators~\cite{huang_pan, huang}. These developments motivate further theoretical frameworks aimed at capturing dynamic light-matter interactions and their role in non-local parity manipulation.

The most established approach to implementing anyon braiding in topological systems relies on multi-wire 2D networks using T-junction~\cite{alicea, aasen, clarke} or Y-junction geometries~\cite{Pandey2023, Harper_PRR19, Khanna_PRB22,HEGDE2020}. These schemes typically involve dynamically tuning the coupling between spatially separated Majorana fermions, a process that can be technically demanding both experimentally and theoretically. In contrast, the potential for realizing fusion and braiding protocols through purely geometric or quantum state-based mechanisms—with no explicit spatial motion—has received relatively little attention.

In this work, we propose an alternative approach to conventional multi-wire architectures by investigating the properties of a single Kitaev chain locally coupled to a quantized bosonic mode within the framework of cavity quantum electrodynamics. Our aim is to explore Majorana fusion and non-Abelian features through photon-mediated interactions in one-dimensional systems, eliminating the need for multi-dimensional braiding geometries. We specifically focus on the ``sweet spot" regime, which—despite lacking full topological protection—has proven experimentally suitable for generating Majorana-like excitations in minimal quantum-dot-superconductor configurations~\cite{Sau2012, Leijnse_PRB12, Dvir2023, tenHaaf2024, Zatelli2024, Bordi_PRL24}. These excitations, often called ``poor man's Majorana modes" (PMMMs), retain non-locality and parity coherence, providing a controlled setting for probing Majorana physics in tunable platforms.

We demonstrate that coupling a minimal Kitaev chain to a single-mode quantum cavity in the ultrastrong coupling (USC) regime enables photon-driven manipulation of non-local fermionic parity sectors. The cavity acts as a dynamical ``quantum scissor"~\cite{mendez23}, effectively splitting the chain and releasing PMMMs into the bulk. This mechanism enables simulation of anyon fusion, emergence of parity observables, and formation of entangled Majorana–Schrödinger cat states.

Crucially, unlike previous works that treated MZMs as static inputs to a cavity system~\cite{contamin, trif1, trif2, Olesia_PRB24}, our approach treats the coupled light-matter system as a unified polaritonic entity. This viewpoint reveals Berry-phase-like signatures and parity control under adiabatic modulation of light–matter coupling, which can be engineered using superconducting circuits or hybrid semiconductor-superconductor devices~\cite{fj, schlawin, mendez20}.

Altogether, our findings establish a photonic framework to emulate and manipulate Majorana fusion physics in one-dimensional systems, broadening the landscape of quantum simulations of non-Abelian anyons and paving the way for light-assisted quantum operations in topologically inspired settings.

As we will demonstrate below, for low and intermediate coupling strengths the polariton state is an entangled matter-light state. Nevertheless, in the ultra-strong coupling regime, the asymptotic decoupling of light and matter degrees of freedom~\cite{prl14,ashida} enables both anyon fusing and braiding processes, mediated by photons, to occur in a KC-cavity system. Additionally, our proposal gives access to highly non-classical Schrödinger cat photon states, supported by the underlying nonlocal Majorana structure. Such features turn out to be within reach in the USC regime due to the burgeoning interest on new cavity quantum electrodynamics (cQED) architectures and material supports~\cite{bloch, fj, schlawin}. The USC is particularly interesting for going beyond such first-order effects. The transition between different photon coherent states goes with corresponding modifications of fermion correlations. Remarkably, in the asymptotic decoupling regime, cavity decoherence becomes harmless to MZM dynamics, allowing non-destructive readout of topological qubits. This provides a distinct advantage over existing QD-based approaches~\cite{Baldelli22,Appugliese2022}. Thus, achieving both the sweet-spot in Kitaev chains and the USC regime in cQED setups opens new routes for probing MZM statistics via light.

This paper is organized as follows: Section~\ref{TheoFrame1} briefly discusses the KC-(local) cavity model and its transformation to Rabi-like models in the so-called sweet-spot parameter regime as considered in this study as well as its theoretic background. In Section~\ref{Sec3}, we discuss the anyon fusion rule for a single site cavity.  In Section~\ref{Sec4}, we present and discuss the braiding protocol for a two-site cavity. Next, in Section~\ref{Sec5} Majorana-Schr\"odinger cat states are addressed. In Section~\ref{Sec6}, we summarize our findings, discuss potential future directions, and explore the possibilities for experimental validation. Further technical details have been collected in~\ref{Appen_A} and~\ref{Apped_B}.

\section{Proposal for a photon assisted testbed of non-Abelian anyon properties}\label{TheoFrame1}

We consider a hybrid light-matter system consisting of a single-mode quantum cavity coupled to a Kitaev chain. The system is described by the Hamiltonian:
\begin{equation}
\hat{H}=\hat{H}_{\rm C} + \hat{H}_{\rm KC} + \hat{H}_{\rm X}.
\end{equation}
Where, the single mode microcavity Hamiltonian takes the usual form $\hat{H}_{\rm C}=\omega \hat{a}^\dagger\hat{a}$, where $\omega$ is the photon quantum energy ($\hbar=1$), with $\hat{a}$($\hat{a}^\dagger$) the annihilation (creation) microwave photon operator. The Hamiltonian $\hat{\mathcal{H}}_{\rm KC}$ describing an isolated Kitaev chain with open boundaries takes the form \cite{kitaev}
\begin{equation}
\hat{H}_{\rm KC} =  -\frac{\mu}{2} \sum_{j=1}^L 2 (\hat{c}_j^\dagger \hat{c}_j - \hat{1})  -  \sum_{j=1}^{L-1} t(\hat{c}_j^\dagger \hat{c}_{j+1} + \hat{c}_{j+1}^\dagger \hat{c}_j) 
 - \Delta (\hat{c}_j \hat{c}_{j+1} + \hat{c}_{j+1}^\dagger \hat{c}_j^\dagger).
\end{equation}
Here, $\hat{c}_j$ and $\hat{c}_j^\dagger$ denote the fermionic annihilation and creation operators at lattice site $j = 1, \ldots, L$, respectively. The parameter $\mu$ represents the chemical potential, $t$ is the nearest-neighbor hopping amplitude, and $\Delta$ characterizes the superconducting pairing strength between adjacent sites. This model describes a one-dimensional topological superconductor, featuring a quantum phase transition between topological and trivial (nontopological) phases at $\mu = 2\Delta$, assuming $\Delta = t$.

One-dimensional topological superconductors are predicted to host robust Majorana zero modes~\cite{Lutchyn_PRL10}. However, in realistic implementations, microscopic disorder and finite-size effects can hinder the unambiguous detection and control of these modes~\cite{Dvir2023,Kouwenhoven2024,Prada2020}. An alternative approach that has gained increasing attention involves engineering effective Kitaev chains atom-by-atom~\cite{NadjPerge2014} or through quantum dot arrays in hybrid semiconductor-superconductor systems~\cite{Dvir2023,tenHaaf2025,Bordin2025}.

Notably, even minimal systems—such as a two-site Kitaev chain-can support so-called ``poor man's Majorana modes" (PMMMs) when operated at the so-called ``sweet spot": vanishing chemical potential ($\mu=0$) and equal hopping and superconducting pairing amplitudes ($t = \Delta$)~\cite{Leijnse_PRB12}. While these states lack full topological protection, they retain key features of MZMs, including spatial separation and parity-based coherence~\cite{Dvir2023}.

Guided by both experimental feasibility and analytical tractability, we consider a Kitaev chain in this sweet spot regime as the basis for our photon-assisted architecture. This minimal configuration provides a highly controllable platform for exploring emergent non-Abelian features, especially when interfaced with quantized cavity fields~\cite{liu,tenHaaf2025}. For vanishing chemical potential the Kitaev Hamiltonian takes a diagonal form in terms of new non-local fermion operators $\hat{d}_{j}$ and ${\hat{d}_{j}^{\dagger}}$, i.e. $\hat{H}_{\rm KC}=2\Delta\sum_{j=1}^{N-1}\left[\hat{d}_{j}^{\dagger}\hat{d}_{j}-\frac{1}{2}\right]=i\Delta\sum_{j=1}^{N-1}\hat{\gamma}_{j,2}\hat{\gamma}_{j+1,1}$ (see Ref.~\cite{gomez18} for their definition in terms of both original Kitaev local operators ($\hat{c}_j^\dagger, \hat{c}_j$) or Majorana fermions ($\hat{\gamma}_{j,1},\hat{\gamma}_{j,2}$)). 

Finally, the KC-cavity coupling term is given by
\begin{equation}\label{Eq:m4}
\hat{H}_{X}=\frac{1}{\sqrt{n_{\rm Cav}}}\sum_{j\in {\rm Cav}} \lambda_j\hat{c}_j^\dagger \hat{c}_j\left ( e^{-i\phi_j}\hat{a}^\dagger+\hat{a}e^{i\phi_j} \right ),
\end{equation}
where  $\lambda_je^{-i\phi_j}$ the complex-valued coupling strength between the KC $j$-site and the cavity field, Cav is the set with the site indices interacting with the cavity~\cite{Fuentes:00}. For convenience, we refer to a single Kitaev site coupled to the cavity as $1CK$, and to two Kitaev sites coupled to the cavity as $2CK$ throughout this work.

\subsection{Bulk single site case}\label{single_site} 
\begin{figure}[t!]
\centering
\includegraphics[width=1\linewidth]{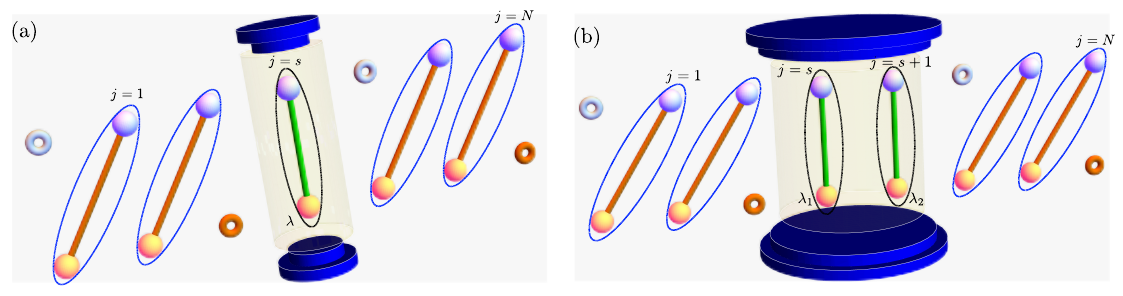}
\caption{\label{Fig_0}  {\bf Schematic representation of Kitaev chain-cavity systems}. In the topological phase, a Kitaev chain is schematically represented by colored balls and links. The balls represent Majorana fermions $\gamma_{j+1,1}$ and $\gamma_{j,2}$ with different colors, indicating the localization of a free MZM as a torus. The cavity is depicted by two blue mirrors, operating as natural topological scissors. The KC-cavity interaction in the USC regime allows the formation of local Dirac fermions located inside the cavity. (a) corresponds to the $1CK$ case while (b) depicts the $2CK$ case. The radiation-matter interaction is represented by $\lambda_r$, with $r=1,2$.}
\end{figure}

We consider the case when the chain-cavity coupling contribution at site $j=s$ of the KC in Eq.~(\ref{Eq:m4}) is the fermion number operator $\hat{c}_s^\dagger \hat{c}_s=\frac{1}{2}\left ( 1+i\hat{\gamma}_{s,1}\hat{\gamma}_{s,2} \right )=\frac{1}{2} \left (1+\hat{d}_{s-1}\hat{d}_s + \hat{d}_{s-1}\hat{d}_s^\dagger-\hat{d}_{s-1}^\dagger\hat{d}_s-\hat{d}_{s-1}^\dagger\hat{d}_s^\dagger \right )$. In panel (a) of Fig.~\ref{Fig_0}, we schematically represent the bulk single-site case. Remarkably, for this coupling term the parity of the number of $d$-fermions is a constant of motion, i.e. a single photon can only produce changes on the $d$-fermion number by $0, \pm2$. Since only two $d$-fermion operators ($\hat{d}_{s-1}$ and $\hat{d}_{s}$) are included in the matter-radiation coupling term, just $4$ matter (KC) states are involved in the full system quantum evolution. Therefore, the Kitaev even state sector is spanned by the basis set $ |-\rangle_z = |{\rm vac}\rangle=| \circ\circ\rangle$,  $|+\rangle_z = \hat{d}_{s-1}^\dagger\hat{d}_{s}^\dagger | \circ\circ\rangle=| \bullet\bullet\rangle$. Where $|{\rm vac}\rangle=| \circ\circ\rangle$ represents the vacuum state, i.e. no $d$-fermion excitations or $| \circ\circ\rangle=0$ for $j=s-1,s$. Consequently, the $1CK$ Hamiltonian reduces to that of an effective Rabi-like model with homodyne rotated quadrature:
\begin{eqnarray}
\hat{H}_1=-2\Delta\hat{\sigma}_z+\omega \hat{a}^\dagger\hat{a}+\frac{\lambda}{2}\left ( e^{-i \phi}\hat{a}^\dagger+e^{i \phi}\hat{a} \right )\left ( 1-\hat{\sigma}_x \right ).
\label{Eq:xz2}
\end{eqnarray} 
We proceed to obtain an approximate energy spectrum in the USC regime where the first term in Eq.~(\ref{Eq:xz2}) is treated as a perturbation. At first order in $\Delta$, two-dimensional subspaces, spanned by basis sets $\{ |-\rangle_x \otimes \hat{D}(\alpha)|n\rangle, |+\rangle_x \otimes |n'\rangle \}$, $n,n'=0,1,2,\ldots$, with \blue{$\alpha=-\frac{\lambda}{\omega} e^{-i\phi}$} and $\hat{D}(\alpha)$ the displacement operator, are separately diagonalized yielding to:
\begin{eqnarray}\label{Eq:zi1}
E_{n,\pm}=\frac{1}{2} \pas{2n\omega-\frac{\lambda^2}{\omega}\pm\sqrt{\frac{\lambda^4}{\omega^2}+16\Delta^2 e^{-|\alpha|^2} L_n^2\pas{|\alpha|^2 }}\; },
\end{eqnarray}
where $L_n(x)$ is a Laguerre polynomial. In the weak-coupling regime ($\lambda < \omega\sim \Delta$) matter states are suited best for exploring the dynamics controlled by $\hat{H}_1$. However, in the USC regime ($\lambda > \omega\sim \Delta$) a better-adapted basis set is provided by occupation number states, $|n_{L,R},n_c\rangle$, of a non-local fermion $f_{L,R}=\frac{1}{2}\left ( \hat{\gamma}_{L}+i\hat{\gamma}_{R}\right )$ and a local fermion inside the cavity, $c_s=\frac{1}{2}\left ( \hat{\gamma}_{s,1}+i\hat{\gamma}_{s,2}\right )$. The basis transformation is:
\begin{eqnarray}\label{Eq:hh2}
|\circ\circ\rangle&=\frac{1}{\sqrt{2}}\left ( |\circ_{L,R},\circ_c\rangle -i |\bullet_{L,R},\bullet_c\rangle\right ), \nonumber\\
|\bullet\bullet\rangle&=\frac{1}{\sqrt{2}}\left (|\circ_{L,R},\circ_c\rangle +i |\bullet_{L,R},\bullet_c\rangle\right ).
\end{eqnarray}
In this regime the ground state is simply expressed as $|GS\rangle \simeq -i|1_{L,R},1_c\rangle\otimes |\alpha\rangle$ with $\alpha=-\frac{\lambda}{\omega}e^{-i\phi}$.

\subsection{Two bulk sites case}\label{Apped_A2} 
The cavity couples with $2$ nearest-neighbor bulk physical sites ($s$ and $s+1$ with $s=2,...,N-2$), or equivalently $3$ bulk bond fermion modes are involved) characterized by the interaction Hamiltonian:
\begin{equation}\label{Eq:m66}
\hat{H}_{X,2}=\frac{1}{2\sqrt{2}}\sum_{r=0,1} \lambda_{r+1} \hat{h}_r \left ( e^{-i\phi_{r+1}}\hat{a}^\dagger+e^{i\phi_{r+1}}\hat{a} \right ),
\end{equation}
where the matter operator $\hat{h}_r$ is given by $1+\hat{d}_{s+r-1}\hat{d}_{s+r} +
\hat{d}_{s+r-1}\hat{d}_{s+r}^\dagger-\hat{d}_{s+r-1}^\dagger\hat{d}_{s+r}-\hat{d}_{s+r-1}^\dagger\hat{d}_{s+r}^\dagger$. Where complex valued $\lambda_1 e^{-i\phi_{1}}$ ($\lambda_2e^{-i\phi_{2}}$)  denote the coupling strength at site $s$ ($s+1$). The total Hamiltonian conserves the parity of the number of fermions, i.e. it can only produce changes on the fermion number by $0, \pm2$. Given the fact that the non-interacting matter-field ground-state has zero $d$-fermions, jointly with zero photons, it is natural to confine this study to the even sector of the fermion Hilbert space. Thus, just $4$ matter (KC) states are involved in the full system quantum evolution. A mapping of the matter states to a two-spin system is thus immediate. In the following, we restrict ourselves to the even KC-sector and $\phi_1=0$. For $\phi_2$, two cases will be addressed, $\phi_2=0$ ($\lambda_2>0$) or $\phi_2=\pi$ ($\lambda_2<0$). In a two-spin basis formed by eigenvectors of $z$-Pauli matrix components $\{ |+,+\rangle_z, |+,-\rangle_z, |-,+\rangle_z, |-,-\rangle_z \}$, the KC states are now associated to:
\begin{eqnarray}
|-,-\rangle_z &=|\circ_{s-1},\circ_s,\circ_{s+1}\rangle, \quad|-,+\rangle_z =|\circ_{s-1},\bullet_s,\bullet_{s+1}\rangle,\nonumber\\
 |+,-\rangle_z&=|\bullet_{s-1},\bullet_s,\circ_{s+1}\rangle,  \quad|+,+\rangle_z =|\bullet_{s-1},\circ_s,\bullet_{s+1}\rangle.\label{Eq:n17}
\end{eqnarray}
The free KC term in the Hamiltonian acquires an Ising-type spin-spin coupling term, yielding finally a two-qubit effective Rabi-like Hamiltonian:

\begin{eqnarray}\label{Eq:xz23}
\hat{H}_2=&-\Delta\left ( \hat{\sigma}_{z,1}+ \hat{\sigma}_{z,2}+\hat{\sigma}_{z,1} \hat{\sigma}_{z,2} \right )+\omega \hat{a}^{\dag}\hat{a}\nonumber\\
&+\frac{1}{2\sqrt{2}}\sum_{r=1}^{2}\lambda_r\left ( e^{-i\phi_r}\hat{a}^\dagger+e^{i\phi_r}\hat{a} \right )\left (1- \hat{\sigma}_{x,r}\right ),
\end{eqnarray}
where $\hat{\sigma}_{\nu,r}$, with $\nu=x,y,z$, denote Pauli matrices for spin $r=1,2$, complex valued $\lambda_{1}e^{-i\phi_{1}}$ and $\lambda_{2}e^{-i\phi_{2}}$ represent the coupling strengths at sites $s$ and $s+1$, respectively (see~\ref{Appen_A}). Restriction to the even fermion parity sector has been kept. Notice that for both $1CK$ and $2CK$ cases, the resonance condition between matter and photonic sub-systems is $\omega=4\Delta$ and the $d$-fermion parity is a constant of motion, i.e. a single photon can only produce changes on the $d$-fermion occupation number by $0, \pm2$.

On the other hand, the convenient and novel Hamiltonian structure, described as a single-qubit or two-qubit Rabi-like model for the $1CK$ and $2CK$ cases, respectively (as given by Eqs.~(\ref{Eq:xz2}) and~(\ref{Eq:xz23})), facilitates both analytical and numerical advancements. In quantum optics, the USC regime of light-matter interactions has been the subject of intense investigation from both theoretical and experimental perspectives~\cite{Frisk_NPR19}. In particular, the role of the diamagnetic term in the USC regime has attracted significant attention~\cite{Tuomas_PRA16,Plenio_PRA16,Nataf2010,Olesia_PRB21} and criticism~\cite{Oliver_PRL11,Luca_PRL12}. For this study, we adopt this framework as a complete and valid description for the USC regime. 

\begin{figure}[t]
\centering
\includegraphics[width=0.6\linewidth]{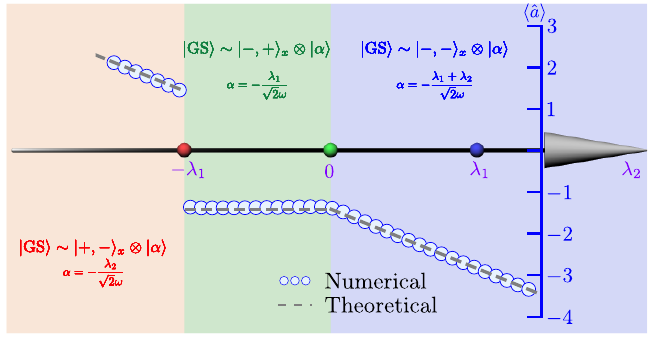}
\caption{\label{Fig_1n}{\bf $2CK$ ground state phase diagram in the USC regime.} Inhomogeneous system, i.e. real valued $\lambda_2 \ne \lambda_1$. Full numerical (open circles) and USC theoretical results (broken line) for the photonic coherent state displacement $\langle\hat{a}\rangle$ are put on top: $\lambda_1/2\Delta=4$ and resonance condition $\omega/2\Delta=2$. See main text for details.}
\end{figure}

In the USC regime, the $2CK$ ground state displays a clear decoupling between matter and photon states for any pair ($\lambda_1,\lambda_2$), as shown in Fig.~\ref{Fig_1n}, except for the special point $\lambda_2=-\lambda_1$ where a superposition of even and odd Schr\"odinger cat states occurs. Simultaneously, a sudden jump of the coherent state displacement $\alpha$ takes place when crossing this special point. Signatures of those transitions are also imprinted in the photon distribution (see Fig.~\ref{Fig_1n}), allowing for optical readout schemes. These special features will be exploited below to set up a braiding protocol.

\section{Fusion rule}\label{Sec3}
MZMs are Ising anyons obeying the fusion rule $\sigma \times \sigma =  1+ \Psi$, where two Ising anyons ($\sigma$) fuse into either a vacuum ($1$) or a standard Dirac fermion ($\Psi$)~\cite{liu, lahtinen}. Fusion rule experiments are in general simpler than braiding experiments, thus they are first-ever candidates for testing the non-Abelian features of MZMs.

To test the non-trivial fusion rule, we focus on the single-site cavity configuration (see Sec.~\ref{single_site}), where only a pair of MZMs is required. Specifically, we implement the fusion protocol in the $1CK$ setup by adiabatically increasing the coupling strength between the Kitaev chain and the cavity. Thus, we start with a single uncoupled KC in its ground state featuring a pair of MZMs located at the chain edges. During the whole process, the total fermionic parity of the KC states is fixed to be even ($+1$) as we start with the ground state of a non-interacting KC-cavity, i.e. $|\chi_1\rangle=|-\rangle_z \otimes |0_{\rm ph}\rangle$ where $|0_{\rm ph}\rangle$ denotes the photon vacuum state. For a KC-cavity strongly coupled system two new MZMs emerge at the neighbor sites of the cavity. 

One immediate consequence of the $H_1$ structure in Eq.~(\ref{Eq:xz2}) is the expression for its USC limit eigenstates: if $\lambda \gtrsim \omega \gg \Delta$, the matter-light eigenstates become uncoupled, given by the separable states $|\Psi\rangle \approx |-\rangle_x \otimes |\alpha\rangle$ where $\hat{\sigma}_x|-\rangle_x =-|-\rangle_x$ and the photon state is nothing but the coherent state $|\alpha\rangle$ with $\alpha=-\lambda/\omega$. Thus, the final matter state, after adiabatically turning on the KC-cavity coupling until the USC regime is reached, turns out to be $|-\rangle_x=\left ( |\circ\circ\rangle-|\bullet\bullet\rangle \right )/\sqrt{2}$ indicating once more that the cavity coupling doesn't alter the total fermionic parity. Therefore, the final matter state is an equal probability superposition of states with no fermions on each side of the cavity and states with a single fermion on each side. The fusion outcome becomes probabilistic, not deterministic, as clearly indicated by the resulting KC-cavity state. This result corresponds to a positive test of the fusion protocol of two Ising anyons. In the case of quantum dot-based Kitaev systems the readout of the fusion outcome can be detected by standard capacitance measurements or tunneling signatures~\cite{Sau:24}. 

\begin{figure}[t!]
\centering
\includegraphics[width=0.7\linewidth]{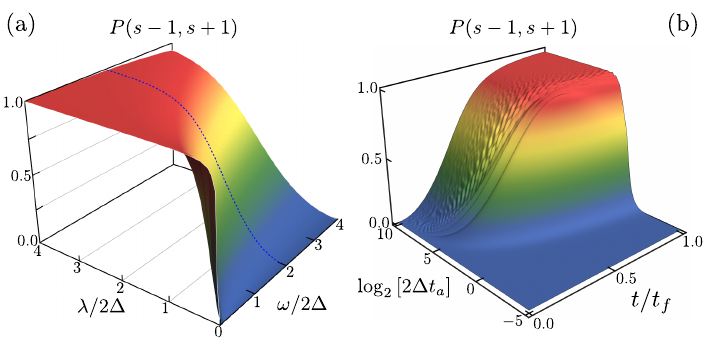}
\caption{\label{Fig_1}  {\bf Fusion rule protocol in a $1CK$ system.}  Panel (a) shows the ground-state fermionic correlation function $P(s-1, s+1)$, where the dashed blue line indicates the resonance condition. In panel (b), we consider the resonant case with a time-dependent linear ramp of the light-matter coupling strength, given by $\lambda(t) = \lambda_{\max} (t / t_a)$, where $t_a$ is the annealing time and $\lambda_{\max} = 4$ is the maximum coupling.}
\end{figure}

In typical experimental platforms, measuring the fusion rule involves evaluating the two-point correlation function or, equivalently, the probabilistic fermion parity~\cite{clarke,liu}. When a single nanowire initially in the topological phase with even fermion parity is split into two halves, the fermion parity readout between the resulting sectors is expected to yield values of $\pm 1$ with a zero average. This outcome reflects the parity combinations allowed by the fusion rules—either even-even or odd-odd—each occurring with an equal probability of $1/2$. To test the validity of the fusion protocol in the $1CK$ system, we numerically simulate the process by diabatically activating the matter-light coupling. Specifically, we compute the fermion correlation function between the two Majorana fermions nearest to the cavity, $P(s-1, s+1) = i\langle \gamma_{s-1,2}, \gamma_{s+1,1} \rangle = -\langle \hat{\sigma}_x \rangle$. This analysis provides a quantitative means to verify the expected behavior of the fusion protocol under the given conditions. Figure~\ref{Fig_1}(a) shows a 3D plot of that correlation in the ground state as a function of photon energy and coupling strength ($\phi=0$). 

\subsection{Time-dependent fusion rule control}
In this subsection, we describe the numerical methods used to obtain the time-dependent results. The light-matter coupling is modeled by a linear ramp $\lambda(t) = \lambda_{\max} \, t / t_a$, where $t_a$ is the annealing time, controlling the ramping speed, and $\lambda_{\max}$ is the maximum coupling strength. At $t = 0$, the cavity is decoupled from the Kitaev chain, and the system is initialized in the ground state of the Hamiltonian given in Eq.~(\ref{Eq:xz2}) with $\lambda = 0$. As the coupling is gradually turned on, the two subsystems begin to interact, governed by the controlled annealing protocol.

The system's wavefunction $\ket{\psi(t)}$ is obtained by numerically solving the time-dependent Schr\"odinger equation. By adjusting the total evolution time $t_f=t_a$, we dynamically explore the transition from the weak-coupling to the ultrastrong-coupling regime. We then compute the fermionic correlation function defined by:

\begin{eqnarray}\label{Correl_1}
P\pap{s-1,s+1}\pap{t_f}&=i\bra{\psi\pap{t_f}}\gamma_{s-1,2}\gamma_{s+1,1}\ket{\psi\pap{t_f}},\nonumber\\
&= - \bra{\psi\pap{t_f}}\hat{\sigma}_x\ket{\psi\pap{t_f}}.
\end{eqnarray}
The correlation function $P(s{-}1,s{+}1)$ serves as a key signature of the fermionic fusion rule. In Fig.~\ref{Fig_1}(b), we present its behavior as a function of the annealing time, which corresponds to the duration over which the matter-photon coupling is gradually turned on. To assess the robustness of the protocol against diabatic errors, we compare this correlation for different ramping rates. Remarkably, we find excellent agreement between analytical predictions and numerical simulations, even for ramping protocols that extend beyond the strict boundaries of the ultrastrong coupling regime—provided the coupling is increased sufficiently slowly.

In summary, what a fusion rule experiment measures is, essentially, a two-point correlation or, equivalently, a probabilistic fermion parity~\cite{clarke,liu}. When a single nanowire, initially in the topological phase with even fermion parity, is split into two, the fermion parity readout between the two new sectors is expected to yield values of 
$\pm 1$ with zero average. This outcome reflects the allowed parity combinations, either even-even or odd-odd, with identical probabilities, $1/2$, and this is precisely what is here found for the $1CK$ case in the USC regime.

\section{Braiding protocol}\label{Sec4}
As a further step, we propose a cavity-chain setup to perform an MZM braiding/exchange process. Most recipes for simulations of braiding protocols in 1D topological systems rely on either Floquet dynamics (Majorana time crystals~\cite{boman}) or quantum dot mediated exchange~\cite{liu, wu}. The former case demands extreme timing control while the latter scheme requires a vanishing dot energy level splitting which hampers an adiabatic evolution harming the goal of a geometric phase $\pi/4$, as required for a perfect exchange between two MZMs. Our goal is to reach the same $\pi/4$-phase by performing an adiabatic cyclic process on the cavity itself, i.e. on the coherent state $|\alpha\rangle$, by leveraging the separable property of the ground state (GS) in the USC regime, i.e. $|{\rm GS}\rangle \simeq  |\chi\rangle\otimes |\alpha\rangle$ with $|\chi\rangle$ a matter KC state. In our setup, the polariton KC-cavity state can be coupled to the Majorana state, and so is not involved in braiding, only initialization (and possibly readout).

\begin{figure}[t!]
\centering
\includegraphics[width=0.8\linewidth]{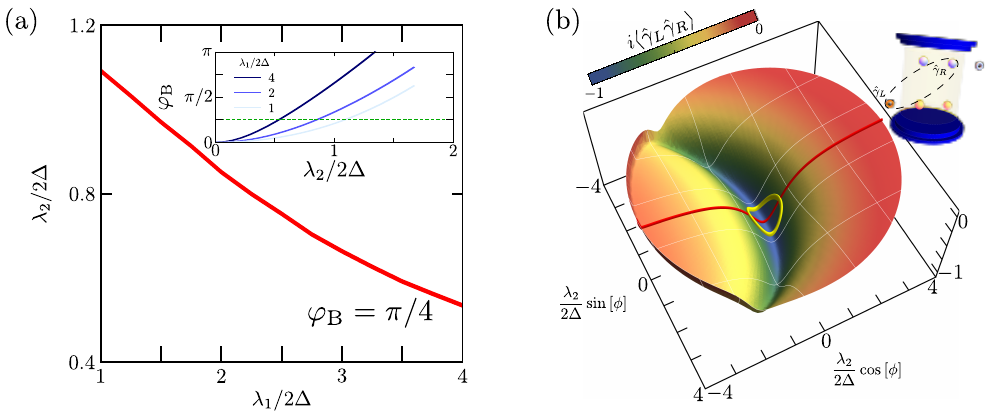}
\caption{\label{Fig_2} {\bf Signature of MZM Braiding in a 2KC Setup.} In panel (a), we depict the numerical relationship between the couplings $\lambda_2$ as a function of $\lambda_1$, where the Berry phase is exactly $\pi/4$. In the inset of panel (a), we show the behavior of the Berry phase by numerical evaluation of Eq.~(\ref{eqberry}) for several values of $\lambda_1$. In panel (b), we illustrate the landscape of parity between $\gamma_{L}$ and $\gamma_{R}$.
}
\end{figure}

To simulate the braiding of two Majorana fermions, $\hat{\gamma}_L$ and $\hat{\gamma}_R$, we consider the $2CK$ setup in which the cavity couples to two adjacent sites in the bulk of the chain: $j = s$ and $j = s+1$. The corresponding light-matter interaction is governed by $H_2$ in Eq.~(\ref{Eq:xz23}), with site-dependent couplings $\lambda_1 e^{i\phi_1}$ and $\lambda_2 e^{i\phi_2}$, respectively. Without loss of generality, we fix $\lambda_1 > 0$ in the ultrastrong coupling regime and set $\phi_1 = 0$, while treating $\lambda_2$ and $\phi_2$ as tunable parameters.\\
\\\
In this configuration, the Majorana fermion just to the left of the cavity, $\hat{\gamma}_L = \hat{\gamma}_{s-1,2}$, consistently behaves as a free MZM. In contrast, the Majorana to the right of the cavity, $\hat{\gamma}_R = \hat{\gamma}_{s+1,1}$, acquires the MZM character only in the limit $\lambda_2 \rightarrow 0$. The braiding protocol we analyze involves these two MFs, $\hat{\gamma}_L$ and $\hat{\gamma}_R$, and is designed to explore how their exchange is mediated through cavity coupling dynamics.\\
\\
For clarity, we schematically indicate which Majorana mode is being manipulated or exchanged at each stage of the protocol using dashed ovals in the inset of Fig.~\ref{Fig_2}. This visual aid helps track the braiding steps and reinforces the interpretation of the cavity as a local control element enabling effective non-Abelian operations within a one-dimensional geometry.

As in the single-site cavity case, the weak-coupling regime of the two-site setup is best described using the matter states defined in Eq.~(\ref{Eq:n17}), which offer a suitable basis for analyzing the dynamics governed by $\hat{H}_2$. However, in the USC regime, a more appropriate representation is given by the occupation number basis $\{ |n_{L,R}, n_{c1}, n_{c2} \rangle \}$, where $f_{L,R} = \frac{1}{2} \left( \hat{\gamma}_L + i\hat{\gamma}_R \right)$ defines a non-local fermion composed of Majorana modes outside the cavity, and $c_1 = \frac{1}{2} \left( \hat{\gamma}_{s,1} + i\hat{\gamma}_{s,2} \right)$ and $c_2 = \frac{1}{2} \left( \hat{\gamma}_{s+1,1} + i\hat{\gamma}_{s+1,2} \right)$ are two local fermions fully contained within the cavity region.\\ 
\\
Further details on this basis transformation and its physical implications are provided in~\ref{Appen_A}. For $\phi_1=0$, the composite ground state in the USC regime becomes
\begin{equation}
|{\rm GS}_{\pm}\rangle\simeq\Bigg\lbrace
\begin{array}{cc}
-i|\Psi_{+}\rangle\otimes |\alpha_+\rangle, &{\rm for }\;\phi_2=0,\\
+i|\Psi_{-}\rangle\otimes |\alpha_-\rangle, &{\rm for }\;\phi_2=\pi.
\end{array}
\end{equation}
Where $|\Psi_{\pm}\rangle=(|\bullet_{L,R},\bullet_{c_1},\circ_{s+1}\rangle\pm|\circ_{L,R},\bullet_{c_1},\bullet_{s+1}\rangle)/\sqrt{2}$, $\alpha_+=-(\lambda_1+\lambda_2)/(\sqrt{2}\omega)$, and $\alpha_-=-\lambda_1/(\sqrt{2}\omega)$ regardless the value of $\lambda_2\le \lambda_1$. In the 1CK limit, i.e. $\lambda_2\rightarrow 0$, the system's ground state turns out to be
\begin{eqnarray}\label{Eq:pq3_main}
|{\rm GS}\rangle &\simeq \frac{1}{\sqrt{2}}\left ( |{\rm GS}_{+}\rangle -|{\rm GS}_{-}\rangle\right ),\nonumber\\
&=|\bullet_{L,R},\bullet_{c},\circ_{s+1}\rangle\otimes |\alpha\rangle.
\end{eqnarray}
With $\alpha=-\lambda_1/(\sqrt{2}\omega)$. The braiding or exchange process between $\hat{\gamma}_L$ and $\hat{\gamma}_R$ MFs, as described by the operator $\hat{U}_B^{(L,R)}=e^{\frac{\pi}{4}\hat{\gamma}_L\hat{\gamma}_R}=\frac{1}{\sqrt{2}}\left ( 1+\hat{\gamma}_L\hat{\gamma}_R\right )$ (performing the exchange transformations $\hat{\gamma}_L\rightarrow \hat{\gamma}_R$ and $\hat{\gamma}_R\rightarrow -\hat{\gamma}_L$), when acting on the ground state produces the result $\hat{U}_B^{(L,R)}|{\rm GS}\rangle=e^{-i\frac{\pi}{4}}|{\rm GS}\rangle$. By using a local phase shifter acting just on the $j = s+1$ chain site~\cite{Shen:17, Arrazola:21, Feofanov:10, Fuentes:00}, a controlled complex coupling strength $\lambda_2 e^{i\phi_2}$ should be possible enabling a cyclic evolution of the cavity coherent state $|\alpha\rangle$ and consequently to access to a specific Berry phase of the whole matter-photon state (dynamic phases which depend on the system's specific time evolution may be experimentally canceled out~\cite{Abdumalikov:13, Berger:12}).

Now, we focus on recovering the same $e^{-i\frac{\pi}{4}}$ global phase factor on the GS by performing an adiabatic cyclic evolution of the coherent cavity field (Berry phase). The GS adiabatic evolution is fully contained in the two-dimensional sub-space spanned by the states $\{ |{\rm GS}_+\rangle, |{\rm GS}_{-}\rangle\}$, as previously assured (for details see~\ref{Apped_B}).
We perform a cyclic evolution of the GS by fixing a $\lambda_2>0$ value while varying the phase $\phi_2=\phi$ of the coupling strength, $0\le \phi\le 2\pi$. Thus, for the GS we have $|{\rm GS}(\phi)\rangle= C_+(\phi) |{\rm GS}_{+}(\phi)\rangle +C_-(\phi)|{\rm GS}_{-}\rangle$.
Notice that the state $|{\rm GS}_{-}\rangle$ does not depend on the phase $\phi$. We proceed with an analytical calculation of the Berry phase using the sub-space-confined evolution of the GS. By direct calculation, we obtain: 
\begin{equation}\label{eqberry}
\varphi_{B}=\int_{0}^{2\pi}\frac{d\phi}{2\omega}\delta E_\phi\pas{1-\frac{\delta E_\phi}{\sqrt{4\Delta^2|\langle\alpha_{-}|\alpha_{+}(\phi)\rangle|^2-\delta E^2_\phi}}}.
\end{equation}
Where, we are using $\delta E_{\phi}=\omega (|\alpha_{+}(\phi)|^2-|\alpha_{-}|^2)$, $\alpha_{-}=-\lambda_1/(\sqrt{2}\omega)$, and $\alpha_{+}(\phi)=-(\lambda_1+\lambda_2 e^{i\phi})/(\sqrt{2}\omega)$. 
If $\Delta\ll\lambda_2$, the Berry phase $\varphi_B\rightarrow \pi \lambda_2^2/\omega^2$, coinciding exactly with the Berry phase of a pure coherent state with $\alpha=-\lambda_2/(\sqrt{2}\omega)$ under an adiabatic cyclic evolution~\cite{Chaturvedi:87}. A numerical evaluation of Eq.~(\ref{eqberry}) allows us to make a plot of $\lambda_2$ as a function of $\lambda_1$ to get a Berry phase of just $\pi/4$, as shown in Fig.~\ref{Fig_2}(a). As it is evident from that figure to reach the appropriate Berry phase the coupling strength with the right KC site $\lambda_2\ll\lambda_1$, validating the ground state structure displayed by Eq.~(\ref{Eq:pq3_main}). Measurement of the photon Berry phase in superconducting cavity systems has been widely discussed previously~\cite{Gasparinetti:16}.

The parity $P_{L,R} =i\langle \hat{\gamma}_L\hat{\gamma}_{R} \rangle$, between $\hat{\gamma}_{L}$ and $\hat{\gamma}_{R}$ MFs, is plotted in Fig.~\ref{Fig_2}(b) as a function of $\lambda_2 e^{i\phi_2}$, for $0\le \lambda_2\le \lambda_1$ and $0\le \phi_2\le 2\pi$. $P_{L,R}\rightarrow -1$ for $\lambda_2 \approx 0$ in agreement with the GS given by Eq.~(\ref{Eq:pq3_main}). The solid red line shows the corresponding results for varying real $\lambda_2$ values from the positive to the negative region. At the same time, the closed curve depicts the $P_{L,R}$ evolution just at the $\lambda_2$ value producing the correct Berry phase of $\pi/4$, emulating the perfect braiding/exchange between $\hat{\gamma}_{L}$ and $\hat{\gamma}_{R}$ MFs. The $\pi/4$ phase emerges from the interplay of geometrical and topological features
under these conditions, with its resilience tied to the robustness of the optical Berry phase readout.

Previous studies employing gate-controlled junctions between topological chains or nanowires have predicted braiding in two-dimensional scenarios, underscoring the need for more accurate modeling of strictly one-dimensional systems. We emphasize that in the present work we address this specific issue in a simpler 1D scenario, providing predictions based on the best available tools to model anyon statistics (fusion rules) and braiding in 1D KC-cavity systems.

\section{Majorana-Schr\"odinger cat states}\label{Sec5}
Photon Schr\"odinger cat states have been intensively studied in circuit quantum electrodynamics scenarios since they offer the opportunity of investigating fundamental tests of quantum theory, for encoding logical qubits and for developing error correction codes~\cite{ayyash}. A Schr\"odinger cat-like state can be written as $|{\rm\bf Cat}_{\pm}\rangle=\pap{|\alpha\rangle\pm|-\alpha\rangle}/N_{\pm}$ with normalization constant $N_{\pm}^2=2\pap{1\pm\exp\pas{-2|\alpha|^2}}$ and sign $+$ ($-$) for even (odd) cat state. Interestingly enough, in the USC regime, a $2CK$ system supports photon Schr\"odinger cat states coupled to Majorana fermion states at the special coupling values $\lambda_1=\lambda_2=\lambda$, $\phi_1=0$, and $\phi_2=\pi$. The $2CK$ ground state becomes:
\begin{eqnarray}\label{Eq:xz3}
 |{\rm GS}\rangle &\simeq |+,-\rangle_x\otimes |-\alpha\rangle+|-,+\rangle_x\otimes |\alpha\rangle,\\
&\simeq | \tilde{\Psi}_{+}\rangle \otimes|{\rm\bf Cat}_{+}\rangle + | \tilde{\Psi}_{-}\rangle \otimes|{\rm\bf Cat}_{-}\rangle.
\end{eqnarray}

\begin{figure}[t!]
\centering
\includegraphics[width=0.5\linewidth]{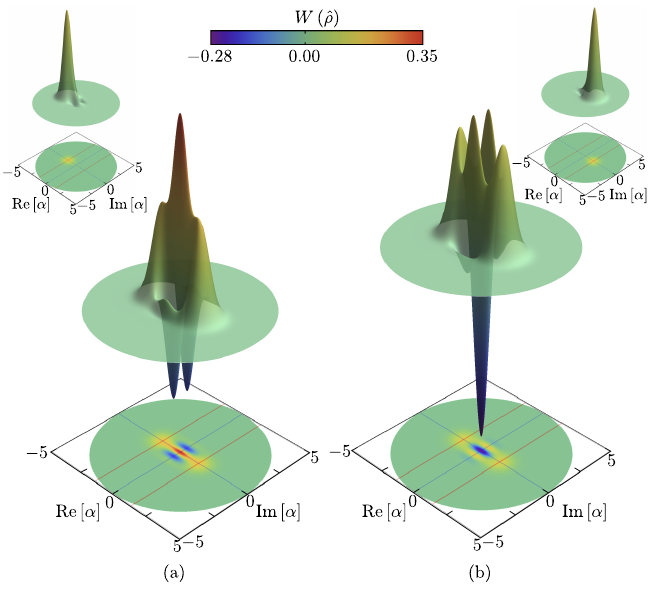}
\caption{\label{Fig_3} {\bf Majorana-Schr\"odinger cat states.} In the main panels (a) and (b), the bosonic Wigner function is represented, conditioned on the symmetry measurement outcome of the fermion state inside the 2CK in the resonance case. An even (panel (a)) or odd (panel (b)) cavity cat state is distilled. The red lines over the density projections correspond to the values of $\alpha = \pm\lambda/(\sqrt{2}\omega)$, with $\lambda = 4$ fixed. Similarly, the insets correspond to the numerical Wigner functions obtained when the ground state is projected onto $\ket{+,-}_x$ and $\ket{-,+}_x$, respectively.}
\end{figure}

Where, we are using $\alpha=-\lambda/(\sqrt{2}\omega)$, $| \tilde{\Psi}_{\pm}\rangle = (|\bullet_{L,R},\circ_{c_1},\bullet_{c_2}\rangle\pm| \bullet_{L,R},\bullet_{c_1},\circ_{c_2}\rangle)/\sqrt{2}$,  and the matter basis set $\{|n_{L,R},n_{c_1},n_{c_2}\rangle\}$ has been used (number occupation representation of the non-local fermion $f_{L,R}$ and two local fermions fully contained inside the cavity, $c_1$ and $c_2$, see~\ref{Apped_A2}). The result in Eq.~(\ref{Eq:xz3}) indicates that the system's ground state is an entangled KC-cavity state: the symmetric combination of the two local fermion ($c_1,c_2$) state couples with an even Schr\"odinger cat state whereas the anti-symmetric two fermion state couples to the odd Schr\"odinger cat state. Therefore, conditioned on the symmetry measurement outcome of the fermion state inside the $2CK$, an even or odd cavity cat state can be distilled. In the resonant case, the ground state of the Hamiltonian given by Eq.~(\ref{Eq:xz23}) is numerically obtained through exact diagonalization from which the reduced photon density matrix immediately is within reach for the GS out of the special point $\lambda_1=\lambda_2=\lambda$, $\phi_1=0$, and $\phi_2=\pi$. As depicted in the upper-lateral plots in Fig.~\ref{Fig_3} the Wigner functions do not display negative features as expected for coherent states. However, for the special set of coupling strengths we are considering in Eq.~(\ref{Eq:xz3}), after a previous projection of the GS on the matter states  $|\tilde{\Psi}_{\pm}\rangle$, the corresponding Wigner function displays negative values in the central plots of Fig.~\ref{Fig_3}, a fingerprint of Majorana induced Schr\"odinger cat states.

The photon cat states arise specifically due to the cavity scissor effect in a $2CK$ system. This effect enables the formation of free, self-adjoint Majorana zero modes within the KC bulk, enabled by the cavity-induced splitting of the chain. These MZMs are responsible for encoding non-local topological qubits, which, in turn, induce the observed photon cat states. This unique interplay between the KC and the cavity could offer a reliable mean to indirectly detect the presence of MZMs through optical signatures. Moreover, mapping the full Hamiltonian into an effective, yet exact, spin-photon model, we dispose of an efficient and exact method to simplify the problem without sacrificing its key physical insights. This Rabi-like mapping has proven to be a powerful tool for extracting non-trivial results, such as the photon cat states, in a clear and computationally manageable framework.

Importantly, the generation of Schr\"odinger cat states in the $2CK$ system is independent of the resonance condition or dispersive condition, and neither a nonlinear Kerr medium is required. Furthermore, the dissipation effect of losing one excitation results, at some random time, in an inversion of parity of the cat state, i.e. even cat states transform to odd cat states (and vice versa).
\section{Summary and discussion}\label{Sec6}

In this work, we addressed a conceptual question: can a single- or double-site cavity embedded in a Kitaev chain effectively extract and control the chain's topological features? By analyzing this setup, we uncovered a rich interplay between cavity quantum electrodynamics and topological quantum matter. We demonstrated that the cavity acts not merely as a passive boundary splitting the chain, but as an active quantum component with its own dynamical degrees of freedom. This makes it essential for generating and mediating non-local quantum correlations between photons and matter.

Our results show that embedding a KC within a spatially selective cavity in the ultrastrong coupling regime enables the realization of non-Abelian anyon properties, such as fusion rules and braiding operations, within a fully one-dimensional platform. These phenomena can be probed through experimentally viable signatures like fermionic parity readouts and photon-induced Berry phases. Additionally, we showed that hybrid Majorana-polariton Schr\"odinger cat states can be generated via the symmetry of fermionic modes interacting within a two-site cavity configuration, opening a novel path for topological quantum information processing.

Importantly, we demonstrated that in the so-called \textit{sweet spot} of the Kitaev chain, often realized in quantum dot arrays, the coupled light-matter Hamiltonian reduces to a Rabi-like model with rotated homodyne quadratures. This mapping simplifies the theoretical treatment and provides a direct link to experimental observables in cavity QED architectures. Furthermore, deviations from the ideal $\mu=0$ condition do not critically affect the non-local character of the Majorana modes, as supported by prior studies~\cite{mendez23}, indicating that fine-tuning requirements can be relaxed.

While our setup exhibits compelling features and supports various non-Abelian signatures, we also acknowledge some of its limitations. In particular, the lack of topological protection in the minimal Kitaev chains ($1CK$ and $2CK$) implies sensitivity to disorder and decoherence, which should be carefully addressed in future experimental implementations. Nevertheless, the ability to control the cavity parameters and the accessibility of measurable observables still make this platform an attractive and realistic testbed for probing Majorana physics.

We also emphasize that our focus has been on proposing a proof-of-principle platform for photon-assisted anyonic phenomena, rather than presenting a fault-tolerant topological quantum computing architecture. As such, our findings should be viewed as a step toward merging cavity QED and topological phases of matter, with potential to evolve into more robust schemes.

Our findings highlight that the proposed cavity-KC platform offers a concrete and scalable route for exploring the quantum control of MZMs using light. In particular, braiding-enabled quantum teleportation protocols become feasible in this architecture, with immediate relevance to current technologies in superconducting circuits and microwave resonators. This positions our model as a practical framework for testing non-Abelian statistics of free Majorana fermions in nanowires or engineered quantum dot arrays.

From an experimental perspective, recent developments in superconducting quantum circuits and quantum dot systems~\cite{Dvir2023,Alvaro} strongly support the feasibility of realizing the proposed setup. KC-like behavior has already been demonstrated in chains of coupled quantum dots, with parameter control sufficient to access the \textit{poor man's} Majorana regime. Coupling these chains to transmission line resonators and applying tailored drives on either the cavity mode or KC sites can allow detection of the predicted phenomena via the transmission spectrum~\cite{trif1,trif2,Arrazola:21,Feofanov:10}, quantum process tomography~\cite{Abdumalikov:13}, or microwave interferometry~\cite{Berger:12}.

Finally, we foresee multiple directions for further research. An intriguing avenue is the generalization to multi-cavity architectures, where geometric phase-based braiding mechanisms could be controlled with high precision. Additionally, exploring the effects of photon losses, dephasing, and other sources of environmental dissipation will be crucial for assessing the robustness of light-induced Majorana processes and for guiding their realization in realistic experimental setups.

\section*{Acknowledgments}
L.Q. and F.J.R. would like to thank Facultad de Ciencias-UniAndes projects: INV-2021-128-2292 and INV-2023-162-2833 for financial support. F.J.G-R and I.A.B-G acknowledge financial support from Spanish MCIN with funding from European Union Next Generation EU (PRTRC17.I1) and Consejeria de Educacion from Junta de Castilla y Leon through QCAYLE project. Additional support from the Department of Education, Junta de Castilla y Leon, and FEDER funds (CLU-2023-1-05) is also acknowledged. C.T. acknowledges financial support from the project “Nanofotonica para Computacion Cuantica” (NanoQuCo) with reference Y2020/TCS-6545 of the Comunidad de Madrid (CAM).
\appendix
\section{Extended properties of two bulk sites case}\label{Appen_A}
 Similarly to the single site cavity case, in the weak-coupling regime matter states given by Eq.~(\ref{Eq:n17}) are suited best for exploring the dynamics controlled by $\hat{H}_2$. Nevertheless, in the USC regime a better-adapted basis set is provided by occupation number states, $\{ |n_{L,R},n_{c1},n_{c2}\rangle\}$, of a non-local fermion $f_{L,R}=\frac{1}{2}\left ( \hat{\gamma}_{L}+i\hat{\gamma}_{R}\right )$ and two local fermions fully contained inside the cavity, $c_1=\frac{1}{2}\left ( \hat{\gamma}_{s,1}+i\hat{\gamma}_{s,2}\right )$ and $c_2=\frac{1}{2}\left ( \hat{\gamma}_{s+1,1}+i\hat{\gamma}_{s+1,2}\right )$. Now the basis transformation is:
\begin{eqnarray}\label{Eq:hh3}
|-,\mp\rangle_z&=\frac{1}{2}\Bigl [ |\circ_{L,R}\rangle \otimes |\bar{\Psi}_{1}^{\pm}\rangle-i |\bullet_{L,R}\rangle \otimes |\bar{\Psi}_{2}^{\pm}\rangle\Bigl ],\nonumber\\
|+,\pm\rangle_z&=\frac{1}{2}\Bigl [ |\circ_{L,R}\rangle \otimes |\bar{\Psi}_{1}^{\pm}\rangle+i |\bullet_{L,R}\rangle \otimes |\bar{\Psi}_{2}^{\pm}\rangle\Bigl ].
\end{eqnarray}
where $|\bar{\Psi}_{1}^{\pm}\rangle =|\circ_{c_1}\circ_{c_2}\rangle\pm|\bullet_{c_1},\bullet_{c_2}\rangle$ and $|\bar{\Psi}_{2}^{\pm}\rangle =|\circ_{c_1},\bullet_{c_2}\rangle\pm|\bullet_{c_1},\circ_{c_2}\rangle$.  In the USC regime, the ground state is simply expressed as $|GS\rangle \simeq |n_{L,R}\rangle\otimes |\Psi_{c_1,c_2, ph}\rangle$ where $|\Psi_{c_1,c_2,ph}\rangle$ is a composite state of fermions inside the cavity $c_1$ and $c_2$, as well as cavity photons. Different regimes of relative values of $\lambda_1$ and $\lambda_2$ should be discriminated. Given $\lambda_1>0$, three different relative coupling sectors should be separately considered, as:

\begin{eqnarray}
\label{GS:r1}|{\rm GS}\rangle &= |\circ_{L,R},\bullet_{c_1},\bullet_{c_2}\rangle\otimes |\alpha_{\rm I}\rangle,\;\quad {\rm for}\; \lambda_2>0, \\
\label{GS:r2}\vert{\rm GS}\rangle &= i|\bullet_{L,R},\circ_{c_1},\bullet_{c_2}\rangle\otimes |\alpha_{\rm II}\rangle, \;\; {\rm for}\; -\lambda_1<\lambda_2<0,\\
\label{GS:r3}|{\rm GS}\rangle &= i|\bullet_{L,R},\bullet_{c_1},\circ_{c_2}\rangle\otimes |\alpha_{\rm III}\rangle, \; {\rm for}\;\lambda_2<-\lambda_1.
\end{eqnarray}
Here, the coherent state displacements are given by $\alpha_{\rm I}=-(\lambda_1+\lambda_2)/(\sqrt{2}\omega)$, $\alpha_{\rm II}=-\lambda_1/(\sqrt{2}\omega)$, and $\alpha_{\rm III}=-\lambda_2/(\sqrt{2}\omega)$. 

\begin{figure*}[t!]
\centering
\includegraphics[width=0.8\linewidth]{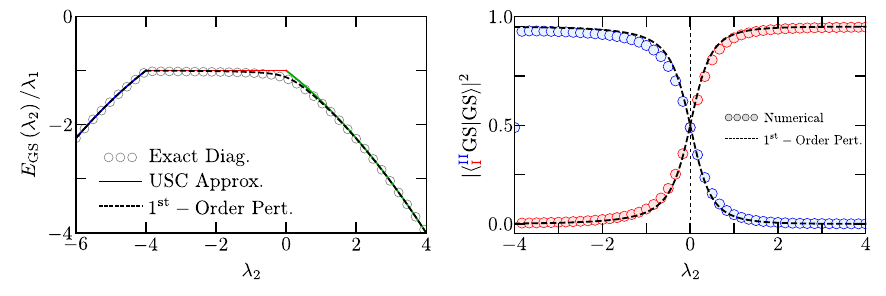}
\caption{\label{Fig_S4} {\bf Ground state indicators for the two bulk sites case.} In the left panel, the symbols represent the ground state energy obtained by exact diagonalization of the Hamiltonian given by Eq.~(\ref{Eq:xz23}) with $\lambda_1 = 4$ for the resonance case. The solid lines represent the USC approximation given by $E_{\rm GS} = -\omega |\alpha|^2$, using $\alpha_{\rm III}$ [Eq.~(\ref{GS:r3})] in the blue region, $\alpha_{\rm II}$ [Eq.~(\ref{GS:r2})] in the red region, and $\alpha_{\rm I}$ [Eq.~(\ref{GS:r2})] in the green region. The dashed black line corresponds to the exact results by first-order perturbation given by Eq.~(\ref{Eq:hh6}). In the right panel, using the same parameters as the left panel, we contrast the numerical results (symbols) and the exact results by employing first-order perturbation theory (dashed lines) for the ground state fidelity. The blue/red symbols correspond to the numerical results obtained using Eq.~(\ref{GS:r1}) and Eq.~(\ref{GS:r2}), respectively. For the exact first-order perturbation results, we used Eq.~(\ref{Eq:hh7}), with the corresponding angle given by Eq.~(\ref{Eq:hh8}).}
\end{figure*}
\begin{figure*}[t!]
\centering
\includegraphics[width=0.8\linewidth]{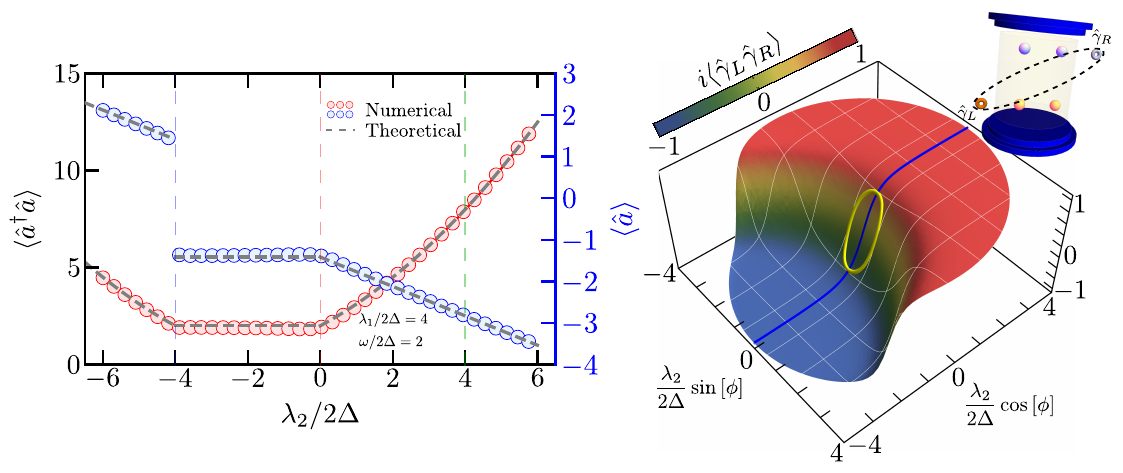}
\caption{\label{Fig_S5} {\bf Parameter Regimes for the Ground State Structure and Braiding in a 2CK Setup.} In the left panel, we present a schematic representation of the ground state structure for 2KC as a function of $\lambda_1$ and $\lambda_2$ [see Eqs.~(\ref{GS:r1}),~(\ref{GS:r2}), and ~(\ref{GS:r3})]. Additionally, we validate the theoretical results for the ground state structure by numerically evaluating the expectation values of the operators $\hat{a}^{\dagger}\hat{a}$ and $\hat{a}$, and contrasting these numerical results with the theoretical predictions. In the right panel, we illustrate the landscape of parity between $\gamma_{L}$ and $\gamma_{R}$. The solid blue line corresponds to the case where $\lambda_2$ is real and changes from positive to negative regimes. The closed yellow ring shows the evolution of the parity, demonstrating the perfect braiding between the Majorana fermions $\gamma_{L}$ and $\gamma_{R}$, but now, in contrast to the main text, we used both outside MFs.
}
\end{figure*}

Numerical results for the ground state energy in each sector are depicted in Figs.~\ref{Fig_S4}-\ref{Fig_S5}, showing a good agreement with predicted USC analytical results, $E_{GS}=-|\alpha|^2\omega$. A crossing behavior is evident at $\lambda_2=-\lambda_1$ due to the quasi-orthogonality of coherent states $|\alpha\rangle$ and $|-\alpha\rangle$ for large values of $\lambda$, whereas an anti-crossing behavior takes place at $\lambda_2=0$. In the latter case, convenient approximate expressions can be found for both energies and ground state composition by considering the free fermion term in the Hamiltonian $\hat{H}_2$, Eq.~(\ref{Eq:xz23}), as a perturbation acting on the sub-space generated by states $\{ |GS_+\rangle,  GS_{-}\rangle\}$. The ground state energy up to the first order in $\Delta$ is given by:
\begin{equation}\label{Eq:hh6}
E_{\rm GS}= -\frac{1}{2}\Bigl [ \omega \left ( \alpha_{\rm I}^2+\alpha_{\rm II}^2 \right ) +
\sqrt{\omega^2 \left ( \alpha_{\rm I}^2-\alpha_{\rm II}^2 \right )^2+4\Delta^2|\langle \alpha_{\rm II}|\alpha_{\rm I}\rangle|^2} \Bigr ],
\end{equation}
while the ground state in the transition regime becomes expressed as the superposition:
\begin{equation}\label{Eq:hh7}
| {\rm GS}\rangle =-\sin\Bigl ( \frac{\theta}{2} \Bigr )| {\rm GS}\rangle_{\rm I}+\cos\Bigl ( \frac{\theta}{2} \Bigr )|{\rm GS}\rangle_{\rm II},
\end{equation}
where
\begin{equation}\label{Eq:hh8}
\tan\left ( \theta \right )=-\frac{4\omega \Delta}{\lambda_2\left ( \lambda_2+2\lambda_1 \right )}e^{-\frac{\lambda_2^2}{4\omega^2}}.
\end{equation}
Thus, it is evident that varying the coupling strength $\lambda_2$ of one physical site with the cavity and crossing the point $\lambda_2= 0$, a cavity allowed {\it flip-flop effect} between the topological fermion state $|n_{L,R}\rangle$ and the local fermion $|n_{c1}\rangle$ is attained.

\section{Braiding protocol in a two-site cavity}\label{Apped_B}
We examine the possibility of reaching the same $e^{-i\frac{\pi}{4}}$ on the GS by performing an adiabatic cyclic evolution of the coherent cavity field (Berry phase). The ground state (GS) adiabatic evolution is fully contained in the two-dimensional sub-space spanned by the states $\{ |{\rm GS}_+\rangle, |{\rm GS}_{-}\rangle\}$, as previously assured. Thus, the effective Hamiltonian matrix in that sub-space becomes:
\begin{equation}\label{Eq:pq8}
\hat{H}_2=\left(\begin{array}{cc}
-\omega| \alpha_+|^2 & \Delta \langle \alpha_-|  \alpha_+ \rangle \\
\Delta \langle \alpha_+|  \alpha_- \rangle & -\omega| \alpha_-|^2 
\end{array}\right).
\end{equation}
With $\alpha_+=-\frac{\lambda_1+\lambda_2e^{i\phi_2}}{\sqrt{2}\omega}$ and $\alpha_-=-\frac{\lambda_1}{\sqrt{2}\omega}$. We perform a cyclic evolution of the GS by fixing a $\lambda_2>0$ value while varying the phase $\phi_2=\phi$ of the coupling strength, $0\le \phi\le 2\pi$. Thus, for the GS we have
\begin{eqnarray}\label{Eq:pq5}
|{\rm GS}(\phi)\rangle= C_+(\phi) |{\rm GS}_{+}(\phi)\rangle +C_-(\phi)|{\rm GS}_{-}\rangle.
\end{eqnarray}
Notice that the state $|{\rm GS}_{-}\rangle$ does not depend on the phase $\phi$. We proceed with an analytical calculation of the Berry phase using Eq.(\ref{Eq:pq8}) and Eq.(\ref{Eq:pq5}). The Berry phase is given by:
\begin{eqnarray}\label{Eq:pq6}
\varphi_B=i\int_0^{2\pi} d\phi \langle {\rm GS}(\phi)| \frac{\partial}{\partial \phi}|{\rm GS}(\phi)\rangle.
\end{eqnarray}
Inserting Eq.~(\ref{Eq:pq5}) into Eq.~(\ref{Eq:pq6}) one gets Eq.~(\ref{eqberry}).

\section*{References}
\bibliographystyle{iopart-num}
\bibliography{Cav_Majo_Bib}
\end{document}